# Application of Microgrids in Addressing Distribution Network Net-Load Ramping

Alireza Majzoobi, Amin Khodaei
Dept. of Electrical and Computer Engineering
University of Denver
Denver, CO, USA
Alireza.Majzoobi@du.edu, Amin.Khodaei@du.edu

*Abstract*— In spite of all advantages of solar energy, its deployment will significantly change the typical electric load profile, thus necessitating a change in traditional distribution grid management practices. In particular, the net load ramping, created as a result of simultaneous solar generation drop and load increase at early evening hours, is one of the major operational issues that needs to be carefully addressed. In this paper, microgrids are utilized to offer a viable and localized solution to this challenge while removing the need for costly investments by the electric utility. In this regard, first the microgrid ramping capability is determined via a min-max optimization, and second, the microgrid optimal scheduling model is developed to coordinate the microgrid net load with the distribution grid net load for addressing the ramping issue. Numerical simulations on a test distribution feeder with one microgrid exhibit the effectiveness of the proposed model.

## Nomenclature

*Indices:*

| | |
|---|---|
| $c$ | Customer connected loads. |
| $ch$ | Superscript for energy storage system charging mode. |
| $dch$ | Superscript for energy storage system discharging mode. |
| $d$ | Index for loads. |
| $i$ | Index for DERs. |
| $j$ | Index for loads at the distribution grid. |
| $t$ | Index for time. |
| $u$ | Superscript for exchanged power with utility grid. |

*Sets:*

| | |
|---|---|
| D | Set of adjustable loads. |
| G | Set of dispatchable units. |
| S | Set of energy storage systems. |

*Parameters:*

| | |
|---|---|
| $DR$ | Ramp down rate. |
| $DT$ | Minimum down time. |
| $E$ | Load total required energy. |
| $F(.)$ | Generation cost. |
| $MC$ | Minimum charging time. |
| $MD$ | Minimum discharging time. |
| $MU$ | Minimum operating time. |
| $UR$ | Ramp up rate. |
| $UT$ | Minimum up time. |
| $\alpha, \beta$ | Specified start and end times of adjustable loads. |
| $\rho_M$ | Market price. |

*Variables:*

| | |
|---|---|
| $C$ | Energy storage available (stored) energy. |
| $D$ | Load demand. |
| $I$ | Commitment state of the dispatchable unit. |
| $P$ | DER output power. |
| $P_M$ | Main grid power. |
| $R$ | Ramping capability of microgrid. |
| $SD$ | Shut down cost. |
| $SU$ | Startup cost. |
| $T^{ch}$ | Number of successive charging hours. |
| $T^{dch}$ | Number of successive discharging hours. |
| $T^{on}$ | Number of successive ON hours. |
| $T^{off}$ | Number of successive OFF hours. |
| $\tau$ | Time period. |
| $u$ | Energy storage system discharging state. |
| $v$ | Energy storage system charging state. |
| $z$ | Adjustable load state (1 when operating, 0 otherwise). |

## I. Introduction

THE evolution of renewable energy over the past few decades has surpassed all expectations, due to significant advantages that they offer, such as reduced operation cost, air pollution reduction, and benefiting from the ubiquitous source of energy. Total worldwide renewable power capacity (excluding large hydro) has been dramatically increased from 85 GW in 2004 to 560 GW by the end of 2013 [1]. However despite the benefits, renewable energy resources challenge the traditional grid management practices, thus their likely impacts on the grid should be also considered. For instance, rapid growth of solar energy as one of the most favorable distributed generation technologies adopted by end-use customers, has changed the typical daily demand curves. A typical daily demand curve rises in the morning and peaks in the afternoon, (especially in the summer as air conditioners are extensively used) and it hits a second highest peak in the early evening. The solar energy resources, however, usually generate the highest amount of power at the noontime and decrease toward sunset, hence they offer the capability of supplying the around-noon power demand but have a marginal effect on early evening peaks. Therefore, rapid growth of solar energy has led to changing traditional afternoon peaks to afternoon valleys which are followed by a steep and problematic peak in early evening hours [2, 3].

In 2013, the California Independent System Operator (CAISO) published a chart depicting the predicted demand curve and potential for "over-generation" occurring at increased penetration of solar energy (Fig. 1). The introduced demand curve by CAISO, also called "duck curve", depicts the potential of solar energy to provide more energy than what can be used by the system in the early afternoon and a severe ramp up in the early evening. This over generation and severe ramp-up in the revised demand curve would be a pressing issue for the utility companies as they may require additional fast response generation units to respond quickly to this change. This ramping effect becomes more severe as the solar energy penetration increases in the power system. As the figure shows the belly of the duck, where solar generation is at a maximum, grows with deployment of solar energy between 2012 and 2020. It is worthwhile to mention that it is planned to supply 20% of the U.S. power consumption by solar energy until 2030.

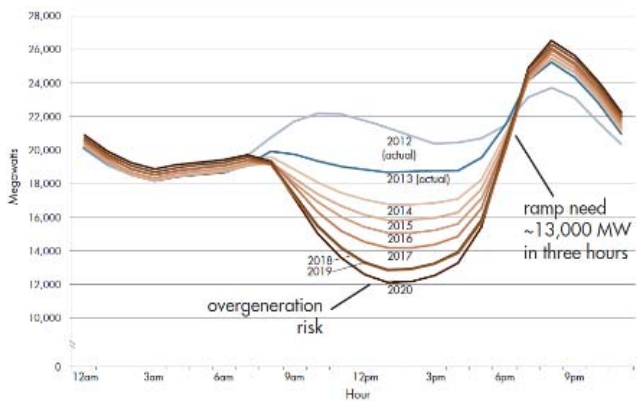

Fig. 1 The current and future estimates of over-generation and ramping effect in California [2].

Addressing the variability of renewable generation has long been an attractive area of research to complement renewable generation forecasting efforts [4, 5]. Uncertainty considerations in power system operation and planning have also significantly increased in the past few years as a large amount of uncertainty sources are integrated in power systems as a result of renewable generation proliferation. The renewable generation coordination problem can be investigated under two contexts of large-scale (which attempts to manage the generation of wind and solar farms) and small-scale (which deals with renewable generation in the distribution level). Examples of large-scale renewable coordination can be found in [6]-[13], which mainly focus on utilizing fast response thermal units, energy storage, and plug-in electric vehicles. Small-scale coordination approaches, on the other hand, mainly focus on various methods of demand side management, such as demand response and demand dispatch [14, 15].

Generation curtailment has been cited as a relatively simple technical solution for over-generation, which occurs with decreasing output power of wind or solar power plants below normal generation by system operators. For wind plants, generation is curtailed by changing the energy captured from the wind by controlling the wind turbines [16]. For solar, this is performed by reducing output from the inverter or disconnecting the plant which requires a specific control system. This control system is available for large renewable power plants but is not economical for small ones. However, although the generation curtailment is a simple technical solution, it would not be a reasonable approach as it reduces the economic and environmental benefits of costly renewable energy deployments [2]. Orienting solar panels to the west-southwest increases the output during the afternoon hours, while reducing output during morning hours [17]. This method, however, is not applicable to large-scale solar plants as they are mainly installed with a tracking system to follow the sun, and also not applicable to small-scale distributed solar plants as they have limitations in terms of orienting the solar panel (for example rooftop solar panels will have fixed orientation). Utilizing energy efficient equipment could also partially mitigate the sharp ramp of early evening hours. As residential lighting accounts for the largest part of the loads in early evenings, using higher-efficiency LED lighting can slightly reduce the early evening peak of the demand curve [17]. In more general settings, demand response can also be considered as a viable solution. However customer-based solutions would require application of smart building management systems, as well as customers' willingness to contribute to solve this issue. Energy storage can be considered as an alternative solution for this problem, however they are still not economically viable.

In this paper, the microgrid is considered as a viable solution for changing the demand curve and mitigating the ramping effects in distribution grids. Microgrids provide a collection of dispatchable generation units, energy storage, and demand response assets, and more importantly, a master controller that can coordinate all these assets while communicating with the electric utility regarding required technical and financial considerations. Leveraging microgrids for addressing renewable generation challenges, as proposed in this paper, will offer a potentially more viable solution in distribution networks, and thus calls for additional studies. The microgrid, as defined by the U.S. Department of Energy, "is a group of interconnected loads and distributed energy resources (DER) within clearly defined electrical boundaries that acts as a single controllable entity with respect to the grid and can connect and disconnect from the grid to enable it to operate in both grid-connected or island-mode" [18]. The microgrid has been introduced as an alternative solution for traditional centralized power generation and bulk transmission. Microgrids offer a localized power generation, control and consumption with remarkable advantages for electricity consumers and the power system. The privileges of microgrids are including but not limited to, power quality improvement, enhanced reliability and resiliency, reduce emissions, network congestion reduction, higher efficiency by decreasing losses, and potential system economics improvement. In grid-connected mode, which is the default mode of microgrids, the microgrid exchanges power with the utility grid to achieve the least-cost supply schedule of local load (an economic operation). However it switches to the island-mode in the event of faults or disturbances in upstream networks to achieve the least load curtailment (a reliable operation) [19-26]. The grid-connected operation mode of the microgrid is considered in this paper, as the microgrid can manage its power exchange with the utility grid to mitigate the severe ramping, and to ensure that the power seen by the utility has manageable ramps.

The rest of the paper is organized as follows. Section II describes the microgrid components, model and relevant equations which used in this paper. The explained model in this section consists of three parts as microgrid components modeling, microgrid ramping capability calculation, and microgrid optimal scheduling formulation. Section III presents numerical examples to show the proposed model applied to a microgrid and discussion on the results of examples and features of the proposed model. Finally, the conclusions are presented in Section V.

## II. MODEL OUTLINE

The first step in modeling the microgrid ramping is to accurately model the microgrid components that provide this ramping. The component modeling is used in two consecutive steps: (1) determining the maximum generation ramping capability of the microgrid, and (2) developing the model for microgrid optimal scheduling to address ramping.

### A. Microgrid Components

The modeled components, which should be considered for ramping studies, consist of local generation units, energy storage, and adjustable/fixed loads, as formulated in (1)-(15).

$$P_i^{\min} I_{it} \leq P_{it} \leq P_i^{\max} I_{it} \quad \forall i \in G, \forall t \quad (1)$$

$$P_{it} - P_{i(t-1)} \leq UR_i \quad \forall i \in G, \forall t \quad (2)$$

$$P_{i(t-1)} - P_{it} \leq DR_i \quad \forall i \in G, \forall t \quad (3)$$

$$T_i^{on} \geq UT_i (I_{it} - I_{i(t-1)}) \quad \forall i \in G, \forall t \quad (4)$$

$$T_i^{off} \geq DT_i (I_{i(t-1)} - I_{it}) \quad \forall i \in G, \forall t \quad (5)$$

$$P_{it} \leq P_{it}^{dch,\max} u_{it} - P_{it}^{ch,\min} v_{it} \quad \forall i \in S, \forall t \quad (6)$$

$$P_{it} \geq P_{it}^{dch,\min} u_{it} - P_{it}^{ch,\max} v_{it} \quad \forall i \in S, \forall t \quad (7)$$

$$u_{it} + v_{it} \leq 1 \quad \forall i \in S, \forall t \quad (8)$$

$$C_{it} = C_{i(t-1)} - P_{it} u_{it} \tau / \eta_i - P_{it} v_{it} \tau \quad \forall i \in S, \forall t \quad (9)$$

$$C_i^{\min} \leq C_{it} \leq C_i^{\max} \quad \forall i \in S, \forall t \quad (10)$$

$$T_{it}^{ch} \geq MC_i (u_{it} - u_{i(t-1)}) \quad \forall i \in S, \forall t \quad (11)$$

$$T_{it}^{dch} \geq MD_i (v_{it} - v_{i(t-1)}) \quad \forall i \in S, \forall t \quad (12)$$

$$D_{dt}^{\min} z_{dt} \leq D_{dt} \leq D_{dt}^{\max} z_{dt} \quad \forall d \in D, \forall t \quad (13)$$

$$T_d^{on} \geq MU_d (z_{dt} - z_{d(t-1)}) \quad \forall d \in D, \forall t \quad (14)$$

$$\sum_{t \in [\alpha_d, \beta_d]} D_{dt} = E_d \quad \forall d \in D. \quad (15)$$

Generating units in a microgrid are either dispatchable or non-dispatchable units. Dispatchable units, such as combined heat and power (CHP), can be controlled by the microgrid master controller, while non-dispatchable units cannot be controlled due to their uncontrollable input sources. The maximum and minimum generation capacity of dispatchable units is formulated in (1), where $I$ represents the unit commitment state (1 when the unit is committed and 0 otherwise). Associated technical constraints are further formulated, including ramp up/down constraints (2) and (3), and minimum up/down time limits (4) and (5). Energy storage has a key role in both grid-connected operation (for energy arbitrage) and islanded operation (for reliability assurance) of microgrids. The minimum and maximum limits of the energy storage charging and discharging are defined in (6) and (7), based on the respective mode. Charging state variable $v$ (1 when charging and 0 otherwise), and discharging state variable $u$ (1 when discharging and 0 otherwise) are used to determine the energy storage operation mode (8). The energy storage is further subject to stored energy amount (9) and capacity (10), determined based on the amount of charged/discharged power, as well as the minimum charging/discharging times (11) and (12). Adjustable loads are subject to minimum and maximum rated powers (13), where $z$ represents loads' scheduling state (1 when load is consuming and 0 otherwise), minimum operating time (14), and the required energy to complete an operating cycle in time intervals (15) [20, 21].

### B. Microgrid Ramping Capability Calculations

The microgrid ramping capability is determined using the following model:

$$R = \min_{R_t} \max_{P,D} |P_{Mt} - P_{M(t-1)}| \quad (16)$$

subject to (1)-(15), and

$$\sum_i P_{it} + P_{M,t} = \sum_d D_{dt} \quad \forall t \quad (17)$$

$$-P_M^{\max} \leq P_{M,t} \leq P_M^{\max} \quad \forall t \quad (18)$$

The inner maximization of (16) calculates the maximum possible amount of microgrid ramping in 24 time intervals of one day. The outer minimization function, selects the minimum amount of microgrid ramping amongst all maximum calculated amounts, hence representing a worst-case that ensures the microgrid can provide at least this amount of ramping in every time interval during a day. This objective is subject to component modeling constraints (1)-(15), as well as the load balance constraint (17) and the utility grid transfer limit (18). The load balance constraint ensures that the sum of power generated by DERs (i.e., dispatchable, non-dispatchable units and energy storage systems) and the power from the utility grid matches the hourly load.

The solution of this model will be the microgrid ramping capability in addressing net load ramping in the distribution grid.

*C. Microgrid Optimal Scheduling Formulation*

The proposed microgrid optimal scheduling for coordinating net load ramping is as follows:

$$Min \sum_t \sum_{i \in G} [F_i(P_{it})I_{it} + SU_{it} + SD_{it}] + \sum_t \rho_{M,t} P_{M,t} \quad (19)$$

subject to (1)-(15), and

$$|P_{M,t-1} - P_{M,t}| \leq R \qquad \forall t. \quad (20)$$

The objective (19) minimizes the microgrid operation cost (including the DER operation cost and the cost of energy purchase from the utility grid), and is subject to component modeling constraints (1)-(15), as well as the calculated ramping capability (20). This problem, however, is further subject to one important constraint to limit the distribution grid net load ramping. Assuming $P^c_{jt}$ as the net load of customers in the same distribution feeder as the microgrid, the total power transferred by the utility to the feeder can be calculated as in (21) and restricted as in (22).

$$P^u_t = P_{M,t} + \sum_j P^c_{jt} \qquad \forall t, j \quad (21)$$

$$|P^u_t - P^u_{t-1}| \leq \Delta_t \qquad \forall t \quad (22)$$

Replacing the value of the utility grid power transfer in (22) by its exact value in (21), and accordingly rearranging the terms based on the microgrid power transfer, (23) will be obtained and added to the problem, in which limits are obtained using (24) and (25).

$$\Delta^{low}_t \leq P_{M,t} - P_{M,t-1} \leq \Delta^{up}_t \qquad \forall t \quad (23)$$

$$\Delta^{low}_t = -\Delta_t - (\sum_j P^c_{jt} - \sum_j P^c_{j(t-1)}) \qquad \forall t \quad (24)$$

$$\Delta^{up}_t = \Delta_t - (\sum_j P^c_{jt} - \sum_j P^c_{j(t-1)}) \qquad \forall t. \quad (25)$$

The utility-imposed constraint (23) ensures that the desirable amount of ramping, i.e., $\Delta_t$ is achieved by proper coordination of the microgrid resources.

### III. NUMERICAL EXAMPLES

A microgrid with four dispatchable units, two nondispatchable units including wind and solar, one energy storage, and five adjustable loads is used for studying the performance of the proposed models. The details of DERs and loads as well as hourly market price are available in [20]. The developed mixed-integer programming problems are solved using CPLEX 12.6.

At the first step the microgrid generation ramping capability is calculated. Fig. 2 depicts the ramping capability (MW/h) and operation cost of microgrid with regards to capacity limit of the line between microgrid and the utility grid. This figure illustrates that the ramping capability of microgrid increases from about 2 MW/h and is saturated at about 11 MW/h as the line capacity increases. Furthermore, the microgrid operation cost is dropped from $12,773 at 2 MW line capacity to $11,660 at 15 MW line capacity. This result advocates that tighter transfer limits will reduce the microgrid capability to provide ramping, while increasing its operation cost. However, there is a saturation point for the microgrid ramping capability in which after that the microgrid ramping will not change.

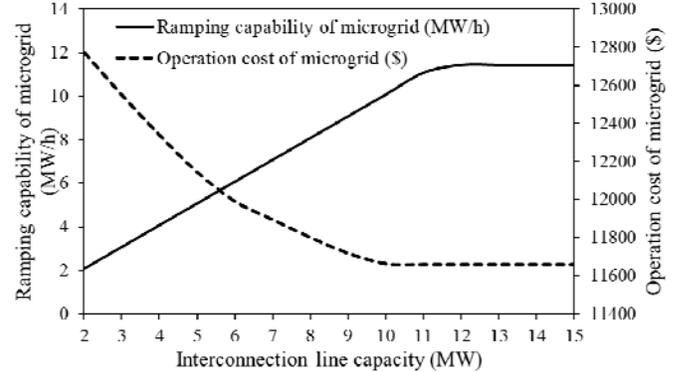

Fig. 2 Ramping capability (MW/h) and operation cost of microgrid for different capacity of interconnection line.

According to the results of this part, the microgrid has ramping capability of 10.079 MW/h, for the case of 10 MW interconnection line capacity, which is considered as a constant in right side hand of constraint (20) for microgrid optimal operation cost analysis. A duck curve is created to show the performance of the microgrid in coordinating the net load ramp. It is assumed that the net load is increased about 15 MW between hours 16 and 19, as well as about 7 MW in just one hour, between hours 18 and 19. This net load is applied to the proposed problem and accordingly Fig. 3 is obtained for microgrid operation cost for various values of the imposed ramping limit by the utility. The microgrid operation cost is decreased with increasing utility grid's generation ramping. Obviously, selection of the lower amount of the utility grid's ramping leads to lower investment and operation costs of utility companies. However, the utility companies should pay to the microgrid for compensation of net load ramping.

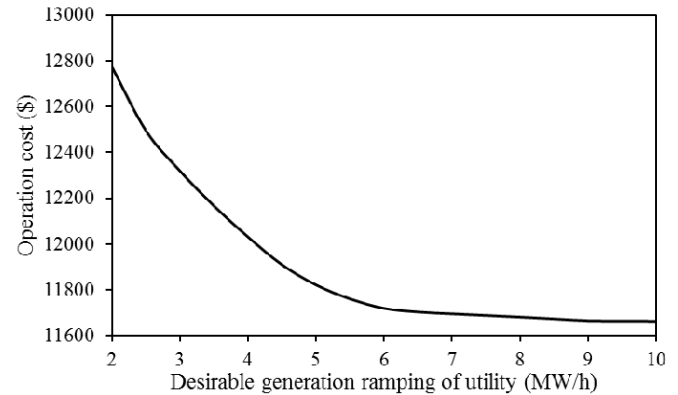

Fig. 3 The microgrid operation cost for various amount of utility grid desired ramping.

This delivered duty to the microgrid leads to higher operation cost which should be paid by utility companies. This would be the main reason of microgrid operation cost decreasing until about 7 MW/h and levelling off at an almost fixed amount after 7 MW/h. For instance, in the case of 8 MW/h ramping, the entire net load ramping is supplied by the utility and the microgrid operation cost will be $11,682. Whereas, in the case of 2 MW/h ramping, 5 MW/h generation ramping should be supplied by the microgrid which leads to a higher operation, equal to $12,773. The operation cost of microgrid without participation in net load ramping addressing, is $11,660. Fig. 4 shows the ramping of utility grid with 2 MW/h and without any constraint, during 24 hours.

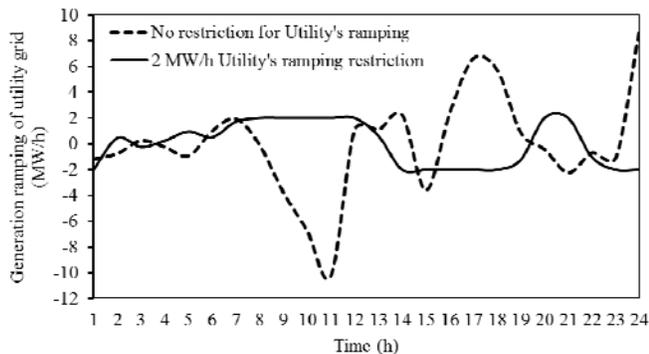

Fig. 4 Generation ramping of utility grid with and without restriction on utility's ramping.

IV. CONCLUSIONS

The microgrid has been used as a solution for mitigation of net load ramping in distribution grids, which occurs due to concurrent decrease in solar generation and increase in consumers' loads. The maximum generation ramping of microgrid was calculated, followed by an optimal scheduling model for coordination of calculated maximum ramping capability with connected loads. The optimization of microgrid operation cost is carried out with consideration of supplying net load's ramping for different amounts of utility grid ramping which further showed the effectiveness of the proposed models.